\documentstyle[12pt]{article}
\begin{document}
\vspace{-2.0cm} 
\begin{flushright}
IFUP-TH 75/96 \\
hep-th/9701083\\
\end{flushright}
\bigskip
\begin{center}
{\large \bf MAGNETIC MONOPOLES, GAUGE INVARIANT DYNAMICAL VARIABLES AND 
GEORGI GLASHOW MODEL\footnote{Partially supported by EC Contract EX-0051 by 
MURST}} 
\end{center}

\bigskip

\begin{center}
\large {Adriano Di Giacomo\footnote{Electronic address: 
digiaco@ipifidpt.difi.unipi.it}  
and Manu Mathur\footnote {Permanent address: S. N. Bose National 
Centre for Basic Sciences, DB-17, Salt \\
\phantom{xxx} Lake, Calcutta  700064, India. 
Electronic address: manu@hpth.difi.unipi.it}} \\
\bigskip
Dipartimento di Fisica dell' Universit\^a and I.N.F.N, \\ 
Piazza Torricelli 2 Pisa, 56100 Italy \\ 
\end{center}

\bigskip
\begin{center} 
{\bf ABSTRACT}\\ 
\end{center}
\bigskip
\noindent We investigate Georgi-Glashow model in terms 
of a set of explicitly SO(3) gauge invariant dynamical variables. 
In the new description a novel compact abelian gauge invariance 
emerges naturally. As a consequence magnetic monopoles occur as 
point like ``defects" in space time. Their non-perturbative contribution to 
the partition function is explicitly included. This procedure 
corresponds to dynamical ``abelian projection" without gauge fixing. In the 
Higgs phase the above abelian invariance is to be identified with  
electromagnetism. We also study the effect of $\theta$ term in 
the above abelian theory.  
\\ 

\newpage 

\begin{center} 
{\bf INTRODUCTION}\\ 
\end{center}

The  subject of magnetic monopoles has been fascinating ever since they 
were proposed by Dirac \cite{dirac1} in 1931  to explain electric charge 
quantization in abelian theory. As a consequence, the abelian gauge group 
becomes compact. 
It has been emphasized in the past \cite{pol1,hooft1} that the global 
nature of the gauge group is important in determining the physical properties 
of a theory.  In pure compact quantum electrodynamics (CQED) on lattice with 
angular gauge fields  and therefore a compact U(1),  the magnetic charges have 
dynamical origin and their condensation is  responsible for the confining 
phase of this theory. Likewise, in non-abelian gauge theories with compact 
gauge groups, magnetic charges can occur naturally because of the hidden 
compact U(1) subgroup(s) \cite{pol1,hooft1}. Their condensation could provide 
a sufficient framework  to  explain quark confinement along the lines of 
dual Meissner effect \cite{hooft2} in CQED. Based on the suggestion 
of 't Hooft \cite{hooft1}, there have been many attempts in the past to get 
effective abelian theories with magnetic monopoles  
via ``abelian projections". In these approaches  certain collective excitations 
of the theory act as an effective ``SU(2) adjoint Higg field" whose zeros are 
the possible locations of the magnetic charges. These issues have been 
largely  addressed at the level of kinematics. However, unlike CQED, there 
has been very little progress at the level of dynamics besides Monte Carlo 
simulations \cite{kron}. The present work is an attempt to understand some of these 
issues along the lines of CQED.  We  will illustrate all the ideas with the 
simplest Georgi Glashow model (GGM) where the Higgs field is one of the 
microscopic fields present in the lagrangian itself. This model has been a 
crucial ground to develop some deep ideas related to non-perturbative dynamics 
in non-abelian gauge theories \cite{hooft3,pol2,pol1,wit2}. In this paper using very 
simple idea of change of basis in the internal space of GGM, we show that   
a] It can be  rewritten completely in terms of explicitly SO(3) gauge 
invariant fields, b] In terms of the above dynamical variables a novel 
{\it compact} abelian gauge invariance ({\it not a subgroup of SO(3)}) 
naturally emerges, c] As a consequence,  the magnetic monopoles in the 
theory now are point like. We explicitly compute their contribution to 
the  partition function. The results presented below are  independent of the 
space time dimension but we illustrate the idea in d=4. The basic idea of this 
work is to define a co-moving 
orthonormal frame or ``body fixed frame" ( BFF) in the internal space with one of 
its axes rigidly attached or identified with the unit  Higgs isovector field 
$\hat{\phi}(x)$. Our motivation to use such frame to describe the dynamics 
 is the following: The other two axes of the BFF being arbitrary up to a local 
rotation around $\hat{\phi}(x)$, the dynamics in the BFF will have a 
built in abelian gauge invariance. Moreover,  by expressing the 
angular motion of the above frame in terms of its ``{\it angular velocities}" we 
will  be able to construct certain SO(3) gauge covariant vector fields  with 
some simple geometrical constraints. Therefore, the dynamics in terms 
of these covariant fields in the BFF will not only be SO(3) gauge invariant 
but will also have the much desired abelian gauge symmetry. 
The above description breaks down at the space time points (``defects") 
where $\vec{\phi}(x) = 0$ and will eventually lead to a non-perturbative
description of magnetic charges. These BFFs have been used in the past \cite{add}  
in the context of sigma model with global O(N) invariance. 
The GGM model is described by the lagrangian  

\begin{eqnarray} 
{\cal{L}}  = {1\over2} D_{\mu}\vec{\phi}.D_{\mu} \vec{\phi} + {1 \over 4}  
\vec{G}_{\mu \nu}(\vec{W}).\vec{G}_{\mu \nu}(\vec{W}) + V(|\vec{\phi}|). 
\label{ggm}
\end{eqnarray}

\noindent Here $D_{\mu}\vec{\phi} \equiv \partial_{\mu}\vec{\phi} - e \vec{W}_{\mu} 
\times \vec{\phi}, \vec{G}_{\mu\nu} \equiv \partial_{\mu}\vec{W}_{\nu} - 
\partial_{\mu}\vec{W}_{\nu} -e \vec{W}_{\mu} \times \vec{W}_{\nu}$.  The potential  
term $V(|\vec{\phi}(x)|)$ will not be crucial in the discussion below. In the case of pure gauge theories the above Higgs fields can be thought 
of as collective excitations of the gluonic fields present in the lagrangian.  
All  the kinematical and most of the dynamical issues discussed below will remain 
unaltered. We describe the isovector Higgs by its gauge invariant magnitude and the 
direction, $\vec{\phi}(x) \equiv \rho(x)\hat{\phi}(x)$. The BFF is specified 
by three orthonormal vectors $\hat{\xi}^{a}(x)$ with $\hat{\xi}^{3}(x) \equiv 
\hat{\phi}(x)$.  The  dynamics of this frame can be described by a set of angular 
velocities defined by 

\begin{eqnarray}
D_{\mu}(\vec{\omega}) \hat{\xi}(x) \equiv 0.  
\label{angvel}
\end{eqnarray} 

\noindent The  ``covariant derivative" $D_{\mu}^{ac}(\vec{\omega}) 
\equiv \delta^{ac}\partial_{\mu} - \epsilon^{abc}\omega_{\mu}^{b}$ 
is defined with respect to body fixed components of the angular velocities.  
Physically, it implies that  any change in $\hat{\xi}^{a}(x)$, (a=1,2,3) 
is perpendicular to itself. We will denote the  body fixed components of 
a vector $\vec{V}$ by latin indices and its components  along $\hat{\xi}^{\pm} 
\equiv \hat{\xi}^{1} \pm i \hat{\xi}^{2}$,  by ${V}^{\mp}$.  
Defining the $U^{\hat{\phi}}(1)$ symmetry transformation 
as $\hat{\xi}^{\pm}(x) \rightarrow exp(\pm i \alpha(x))\hat{\xi}^{\pm}(x)$, 
one  gets the following induced transformations on the angular 
velocities, $\omega_{\mu}^3(x) \rightarrow \omega_{\mu}^3(x) 
+ \partial_{\mu} \alpha(x)$ and $ \omega_{\mu}^{\pm}(x) \rightarrow 
exp(\pm i \alpha(x)) \omega_{\mu}^{\pm}(x)$. 
Thus,  under $U^{\hat{\phi}}(1)$ transformation $(\omega_\mu^\pm(x), 
\omega_\mu^3(x))$ transform like charged matter and abelian gauge field
respectively. In the above description in terms of angular velocities we have 
replaced the unit Higgs vector $\hat{\phi}(x)$ (i.e 2 angular degrees of freedom) 
by a set of 12 angular velocities. Therefore, this new dynamics is highly 
constrained. If we define a field strength tensor corresponding to matter fields 
$\hat{\phi}(x)$ by $ \vec{F}_{\mu \nu}\left(\vec{\omega}(x)\right) \equiv 
\partial_\mu \vec{\omega}_\nu - \partial_\nu \vec{\omega}_\mu + \vec{\omega}_\mu 
\times \vec{\omega}_\nu$, then the constraints are simply 
$\vec{F}_{\mu\nu}\left(\vec{\omega}(x)\right) = 0$. In the space fixed frame (SFF) 
if $\theta(x)$ and $\psi(x)$ are the polar and azimuthal angles of $\hat{\phi}(x)$ 
then the solutions of the above equations are the ``pure gauge"  orthogonal matrices 
$O(\theta,\psi,\alpha)$ describing the relative orientation of the BFF and SFF
in (\ref{angvel}). We thus recover the 
$U^{\hat{\phi}}(1)$ gauge angle along with the 2 
compact degrees of freedoms of ${\hat{\phi}}(x)$  which appeared explicitly 
in the partition function.  Till now we have not described the hidden  
topological aspects of the theory. 
Nonabelian gauge theories also contain topological configurations with 
magnetic charge specified by their topology. Infact, t-Hooft Polyakov 
monopoles in Georgi Glashow model occur as 
stable solitons satisfying the classical equations of motions.  
In the context of abelian gauge theories  with external magnetic charges  or 
vortices \cite{man}, the multivalued angular nature of the matter fields plays 
crucial role in making topological properties manifest at the level of the 
partition function. This is equivalent  to the role played by ``compact photon" 
in CQED. Motivated by these abelian results, we characterise the 
BFF with Euler 
angular co-ordinates of $\vec{\phi}(x)$ with respect to the SFF, i.e,  
$\hat{\phi}(\vec{x})   =  (cos \psi sin \theta,~ sin  \psi sin \theta,
~ cos \theta)$,   
$\hat{\xi}^{1}(\vec{x})  =  (cos \psi cos \theta,~ sin  \psi  cos \theta,
~ -sin \theta)$ and  
$\hat{\xi}^{2}(\vec{x})  =  (- sin \psi, ~cos \psi, ~0)$. 
The corresponding angular velocities in this frame are given by: 
$\omega^{1}_{\mu} = sin(\theta(x)) \partial_{\mu} \psi (x), ~~ \omega^{2}_{\mu} 
= - \partial_{\mu} \theta(x),~~  \omega^{3}_{\mu} 
= - cos (\theta(x)) \partial_{\mu} \psi (x)$. 
We see that $\omega_{\mu}^{3}$ is not defined along the polar 
axis in the internal space where $\theta(x) = 0$ or $\pi$ and around which 
$\psi(x)$ can be multivalued. This is non-abelian analogue of 
decomposing complex abelian Higgs field into its radial and multivalued angular part.  
From now onwards the  space time points where  (${\hat{\phi}} =(0,0,\pm 1)$) with 
multivalued $\psi(x)$ will be called ``singular points". 
Around these points  the third component of the constraint 
$\vec{F}_{\mu\nu}(\vec{\omega}) = 0$ gets modifies by an extra contribution: 
 
\begin{eqnarray}
F_{\mu\nu}^{3({\it{np}})}(\vec{\omega})  =  
- cos\theta{(x)} (\partial_{\mu}\partial_{\nu} - \partial_{\nu}\partial_{\mu}) 
\psi(x) \simeq  \pm (\partial_{\mu}\partial_{\nu} 
- \partial_{\nu}\partial_{\mu}) \psi^{sing}(x).  
\label{cons3}  
\end{eqnarray} 

\noindent In the equation (\ref{cons3}) the superscript $({\it{np}})$ 
stands for ``non-perturbative" and its origin will become clear after we include 
the gauge fields. $\psi^{sing}(x)$ is the multivalued part of $\psi(x)$. 
Uptill now the above results were only regarding the kinematical 
aspects of the Higgs fields. We now extend these kinematical aspects to include 
the gauge fields. The induced $U^{\hat{\phi}}(1)$ transformations on the 
body fixed components of the gauge fields are,  
$W_\mu^{3}(x)  \rightarrow  W_\mu^{3}(x), 
W_{\mu}^{\pm}(x)  \rightarrow  exp(\pm i\alpha(x)) W_{\mu}^{\pm}(x)$.
Under SO(3) gauge transformation $(\vec{\lambda})$ the BFF undergoes 
a rotation $\hat{\xi}^{a} \rightarrow \hat{\xi}^{a} + \vec{\lambda} 
\times \hat{\xi}^{a}$. Therefore, the angular velocities (\ref{angvel}) 
transform as $\vec{\omega}_\mu \rightarrow \vec{\omega}_\mu  
+ \vec{\lambda} \times \vec{\omega}_\mu - \partial_{\mu}\vec{\lambda}$.
At this stage we can straight away define SO(3) covariant 
gauge fields  $\vec{Z}_{\mu}(x)$ and their $U^{\hat{\phi}}(1)$ transformation 
laws  by:  

\begin{eqnarray}
\vec{Z}_\mu(x)  \equiv e\left(\vec{\omega}_\mu(x)  + e \vec{W}_{\mu}(x)\right) ~~~~~~~~~~~~~~~~~~~~~\nonumber \\ 
Z_{\mu}^{\pm}(x) \rightarrow exp(\pm i \alpha(x)) Z_{\mu}^{\pm}(x), ~~
Z_{\mu}^{3}(x) \rightarrow  Z_{\mu}^{3}(x) + {1 \over e} \partial_{\mu} \alpha(x). 
\label{zz}
\end{eqnarray}

\noindent From now on  we will often denote $Z_{\mu}^{3}$ by $A_{\mu}$. 
Under SO(3) gauge transformations the $\vec{Z}_\mu$
transforms like a vector and therefore its components in the BFF are 
explicitly  gauge invariant.   This was one of the motivations to 
describe the dynamics in the moving frame. 
With slight abuse of language we will call $(A_{\mu}(x),Z_{\mu}^{\pm}(x))$ 
as the  ``photon" and the corresponding charged matter fields. 
We will now proceed to define the quantum field theory in terms of these 
SO(3) gauge invariant fields with $U^{\hat{\phi}}(1)$ gauge invariance. 
Using the constraints (\ref{cons3}) and $F^{\pm}_{\mu\nu}(\omega (x)) \simeq
 0$, we find that 

\begin{center} 
$\vec{G}_{\mu\nu}(\vec{W})   =  \left(\partial_\mu 
{Z}_{\nu}^{a}(x) - \partial_\nu {Z}_{\mu}^{a}(x) - e \epsilon^{abc}
{Z}_{\mu}^{b}(x) {Z}_{\nu}^{c}(x)\right) \hat{\xi}^{a}(x) + 
{1 \over e} {\vec{F}}_{\mu\nu}^{(\it{np})}(\psi^{sing})$. 
\end{center} 
\begin{eqnarray}
D_\mu (\vec{W}) \hat{\phi}(x) & = & \epsilon^{\alpha\beta}Z_{\mu}^\alpha(x)
 \hat{\xi}^{\beta}(x) 
\label{hke}
\end{eqnarray}

\noindent Here $(\alpha,\beta)$ vary from 1 to 2. 
At regular points with $\vec{F}^{({\it{np}})}(x) = 0$ the  equations 
(\ref{hke}) are easy to interpret.  Under the SO(3) gauge 
transformation both sides of these equations transform covariantly and 
$\hat{\xi}^{a}(x)$  itself being covariant, ${Z}^{a}_{\mu}(x)$ (a=1,2,3) are 
left invariant. At singular points  or equivalently the non-perturbative sector 
of the theory, only the third component of 
${\vec{F}}_{\mu\nu}^{(\it{np})}(\psi^{sing})$ 
along the $\hat{\phi}$ direction contributes 
to the flux with strength proportional to  $({1 \over e})$. 
This singular flux is given by $\pm {2\pi N \over e}$ 
where $N$ is  the homotopy index of the mapping $S^{1}_{physical} \rightarrow 
S^{1}_{internal}$ provided by the multivalued $\psi^{sing}$ field. 
The Noether current of the symmetry transformation (\ref{zz}) is 
$J^{\hat{\phi}}_{\nu} = 2 e Im \Big(Z_{\mu}^{-}D_{\mu}(A)Z_{\nu}^{+} + 
Z_{\mu}^{+} D_{\nu}(A)Z_{\mu}^{-} - {1 \over 2} \partial_{\mu} 
\left(Z_{\mu}^{+}Z_{\nu}^{-}\right)\Big)$. 
The  $U^{\hat{\phi}}(1)$ covariant derivative $D_{\mu}(A)$ is defined as  
$D_{\mu}(A(x))Z_{\nu}^{\pm} \equiv 
\partial_{\mu} Z_{\nu}^{\pm} \mp i e A_{\mu} Z_{\nu}^{\pm}$. 
$J^{\hat{\phi}}_{\nu}$  is explicitly SO(3) invariant and is independent of 
the initial SO(3) Noether currents. Therefore, the  abelian field strength 
tensor and the corresponding Maxwells equations are given by 

\begin{eqnarray} 
F_{\mu\nu}^{U(1)} = \partial_{\mu} A_{\nu}(x) 
- \partial_{\nu} A_{\mu}(x) - {1 \over e}  F_{\mu\nu}^{3({\it{np}})} \nonumber \\  
\partial_{\mu}F_{\mu\nu}^{U(1)}  =   J^{\hat{\phi}}_{\nu}~~, ~~~~~~~ 
\partial_{\mu}\tilde{F}_{\mu\nu}^{U(1)}  \equiv   K^{mag}_{\nu}. 
\label{maxwell}
\end{eqnarray}

\noindent Here $\tilde{F}_{\mu\nu} \equiv {1 \over 2} \epsilon^{\mu\nu\rho\sigma} 
\tilde{F}_{\rho\sigma}$. The last  equation in (\ref{maxwell}) is the Bianchi identity. 
The magnetic current $ K^{mag}_{\nu} = {1 \over e}  
\partial_{\nu} \tilde{F}^{3({\it{np})}}_{\mu\nu}$ is the topological 
current  which is trivially conserved. 
Note that it is nonzero only at ``defects" where $\vec{\phi}(x) = 0$ and  
is also SO(3) gauge invariant.   The above singular points where 
$F_{\mu\nu}^{3({\it{np}})}$ is non zero form a 2 dimensional manifold (Dirac 
world sheet) which can be characterised by two parameters $(\sigma_{1},\sigma_{2})$ 
and a point on this sheet is given by a four vector 
$X_{\mu}(\sigma_{1},\sigma_{2})$. We will also choose 
$\sigma_{1}$ to describe the time evolution.   
In the above abelian gauge theory $F_{\mu\nu}^{3({\it{np}})}$  should 
be thought of as  ${\it{multivalued}}$  $U^{{\hat\phi}}(1)$ gauge transformation on 
$A_{\mu}$ by $-\psi^{sing} (\theta =0)$ and $+\psi^{sing} (\theta = \pi)$ along the 
trajectories $X_{\mu}(\sigma_{1},\sigma_{2})$ at fixed time. Therefore, these  
correspond to two infinitely thin Dirac string 
with $ \pm ({2\pi N \over e})$ unit of magnetic flux respectively. The magnetic 
monopoles (anti-monopoles) are located at points where these two strings meet or  
$\hat{\phi}$ flips between (0,0,$+$1) and (0,0,$-$1). 
It is clear that the origin of the above Dirac strings is the same as that of the 
Abrikosov Nielsen Olesen vortices in the abelian Higgs model. Moreover, the flipping 
of the extra  ({\it{non-abelian}}) degree of freedom of the Higgs field described by 
its polar angle $\theta(x)$ in  its present angular description provides the above 
strings with ``defects"  which are the gauge invariant locations of the  magnetic 
charges in the theory.  
Infact, this construction of monopoles in non-abelian gauge theory  
is similar to that of in CQED. The  $(F_{\mu\nu}^{U(1)})^{2}$  part of the action 
defined in (\ref{maxwell}) is just the naive continuum limit of the pure CQED action. In the 
following we will extend the above analogies further. 
To recast the magnetic term in a more standard form in the partition function,  
we divide the measure over azimuthal angle into singular and regular parts \cite{man} 
at each space time point  i.e $\int d\psi(x) = \int{d\psi^{reg}}(x) 
\int{d\psi^{sing}}(x)$, here $\psi^{sing}$ is the multivalued part contributing to the 
right hand side of (\ref{cons3}).  
Noting that the  integrations over $\hat{\phi}(x)$ can be replaced by the Haar measure
over the orthogonal matrices $O(\theta(x),\psi(x),\alpha(x))$, the single valued 
angular integration can be traded off with the integration over the angular velocities 
by the following two identities:

\begin{center}
$\int d\vec{\omega}_{\mu}^{a}(x) \delta\left(\vec{\omega}_{\mu}^{a}(x) -
Tr\left(L^{(a)} O(x) \partial_{\mu} O^{-1}(x)\right)\right)  \equiv  1 $ \\ 
$\int dO\left(\theta,\psi,\alpha\right)
\delta \big(\omega_{\mu}^{a} - Tr \left( L^{a} O\partial_{\mu}
O^{-1}\right)\big) 
  =   \int d\psi^{sing}\delta 
\left(F_{\mu\nu}^{a}\left(\vec{\omega}\right) 
+ F_{\mu\nu}^{a{(\it{np})}}(\psi^{sing})\right)$  
\end{center}

\noindent We have suppressed the product over space time points,  
Lorentz and color indeces in the above measures. The action has only 
implicit dependence on $\omega_{\mu}$ through $Z_{\mu}$. Therefore, 
changing the variables in the measure from  $\vec{W}_{\mu}(x)$ to $\vec{Z}_{\mu}(x)$,  
we  get the partition function only in terms of the gauge invariant fields 
$(\rho(x),\vec{Z}_{\mu}(x))$ and the term containing the topological magnetic currents 
of the theory. We will now rewrite this term with Dirac strings.  
It is convinient to describe  the above singularities in a generic 
configuration in terms of  gauge invariant world lines  of the monopoles 
$X^{i}(\sigma_{1})$ and a set of corresponding integers $m^{i}$ describing their 
magnetic charges in the units of ${4\pi \over e}$. A generic monopole
current is  given by 
$K_{\mu}(x) = \sum_{i=1}^{\infty} m^{i} \int d\sigma_{1} {dX^{i}_{\mu}(\sigma_{1})
 \over d\sigma_{1}} \delta^{4}(x-X^{i}(\sigma_{1}))$
Therefore,  the magnetic term in (\ref{maxwell}) is given by  

\begin{eqnarray} 
\tilde{F}_{\mu\nu}^{3({\it{np}})}(x) = \pm{4\pi \over e} \sum_{i=1}^{\infty} m^{i} \int 
d^{2}\sigma \epsilon^{\alpha\beta}\left(\partial_{\alpha}X^{i}_{\mu}(\sigma) 
\partial_{\beta}X^{i}_{\nu}(\sigma)\right) \delta^{4}\left(x-X^{i}(\vec{\sigma})\right). 
\label{dirr}
\end{eqnarray}      

\noindent  This is just the term introduced  by Dirac \cite{dirac} in the 
context of abelian gauge theory where point particles carrying magnetic 
charges were put  by hand. The partition function corresponding to (\ref{ggm}) 
now is given by 

\begin{eqnarray}
Z  & = & \sum_{m_{1},..,m_{\infty}} \prod_{i=1}^{\infty} \int dX_{\mu}^{i}
(\vec{\sigma}) J(X_{\mu}) \int \rho^{2} d\rho\int dZ_{\mu}^a   exp-S\left(
\rho,\vec{Z}_{\mu},X^{i}_{\mu}\right) \nonumber \\ 
S & = & \int\Big[{1 \over 4} \left(F_{\mu\nu}^{U(1)}\right)^{2} 
+ {1 \over 4}\left(D_{\mu}(A)Z_{\nu}^{+} - D_{\nu}(A)Z_{\mu}^{+}\right).h.c 
+  {ie \over 2}F_{\mu\nu}^{U(1)}Z_{\mu}^{+}Z_{\nu}^{-} \nonumber \\
& - & {1 \over 16}\left(Z_{\mu}^{+} Z_{\nu}^{-} - h.c\right)^{2}   
+  {1 \over 2}\left(e^{2} \rho^{2}Z_{\mu}^{+}Z_{\mu}^{-}
 +   \left(\partial_{\mu}\rho\right)^{2}\right) + V(\rho)\Big] d^{4}x  
\label{final}
\end{eqnarray}

\noindent Here J(X) is the Jacobian \cite{emil} due to the change of the measure to 
the string world sheet. Note that the $U^{\hat{\phi}}(1)$ invariance (\ref{zz}) of the  
partition function (\ref{final}) is manifest.  This is an exact result and no 
gauge fixing has been done. In the broken 
phase where $\rho(x)$ = constant + fluctuations,  the 
above partition function describes the interaction of photon with the 
charged massive spin 1 gauge bosons and magnetic monopoles.  
The physical fields here  are explicitly gauge invariant 
and  have the right electric charges under $U^{\hat{\phi}}(1)$. 
It is easy to see that by choosing  the  unitary gauge  
$\hat{\phi}(x) = (0,0,1) \forall x$, possible only   
in the perturbative sector $(m^{i} = 0, \forall i)$, we recover the standard 
results.  At this stage, having an exact abelian theory with magnetic monopoles 
in the partition function, we can also convert them into dyons 
by adding the CP violating $\theta$ term \cite{wit2} in the action (\ref{ggm}):
$\Delta{\cal{L}}  =  \theta {e^{2} \over 32\pi^{2}} \epsilon^{\mu\nu\rho\sigma}
F_{\mu\nu}^{U(1)}(A_{\mu})F_{\rho\sigma}^{U(1)}(A_{\mu})$ 
The equation of motion of $A_{\mu}$ in (\ref{maxwell}) are 
now modified and acquires a $\theta$ dependent term: 
$\partial_{\mu}F_{\mu\nu}^{U(1)}  =   J^{\hat{\phi}}_{\nu} + 
{\theta e^{2} \over 8 \pi^{2}} K^{mag}_{\nu}$.
Therefore, all the magnetic charges with strengths $(m_{1},.....m_{\infty})$ 
also acquire electric charges 
$(q_{1},.....q_{\infty}) = {e \theta \over 2 \pi}(m_{1},.....m_{\infty})$ 
leading to generalized  Schwinger quantization condition.  
The term added above is not a surface term because of the Dirac strings. 
It is also interesting to contrast the above formulation with  some of the 
standard   results in the literature. In the broken phase and in the 
Higgs vacuum: $D_{\mu}(\vec{W})\hat{\phi}(x) \approx 0$. Therefore, 
$e \vec{W}_{\mu}(x) \approx \hat{\phi}(x) \times \partial_{\mu} \hat{\phi}(x) 
+ e \hat{\phi}(x) \tilde{A}_{\mu}(x)$ and $\tilde{A}_{\mu}(x) \equiv 
\hat{\phi}(x).\vec{W}_{\mu}(x)$ is identified with the photon field. Our identification 
of photon differs from this  by the third component of the angular velocity and 
is explicitly gauge invariant. The abelian field strength tensor defined in 
(\ref{maxwell}) is the same as the one proposed in \cite{hooft3}.   
The other proposal in the literature  is simply $\hat{\phi}(x).\vec{G}_{\mu\nu}(x)$. 
However, this will not correspond to (\ref{maxwell}).

Acknowledgement: It is a pleasure to thank Ken Konishi for useful discussions.


\begin{thebibliography}{abcdefghi} 
\bibitem{dirac1} P.A.M. Dirac Proc. Roy. Soc. London  {\bf A 133} (1931) 60 
\bibitem{pol1} A. M. Polyakov, Phys. Lett. {\bf B59} (1975) 82, A. M. Polyakov 
Nuc. Phys.{\bf B 120} (1977) 429
\bibitem{hooft1} G. 't Hooft Nuc. Phys. {\bf B190 [FS3]} (1981) 455. 
\bibitem{hooft2} G. 't Hooft, High energy physics, ed. A. Zichichi 
(Editorice Compositori, Bologna, 1976), S. Mandelstam, Phys. Rep. C {\bf 23} (1976) 245. 
\bibitem{kron} A. S. Kronfeld et. al. Nuc. Phys. {\bf B 293} (1987) 461, For a review 
see:  A. Di. Giacomo, LAT 95, Nucl. Phys. B (Proc. Suppl.) 47 (1996) 136, 
hep-lat/9509036. 
\bibitem{hooft3} G. 't Hooft, Nuc. Phys. {\bf B79} (1974) 276. 
\bibitem{pol2} A. M. Polyakov, JETP Lett. {\bf 20} (1974) 194.  
\bibitem{add} A. D' Adda, M. L\"uscher, P. Di Vecchia Phys. Rep. {\bf 49} 
(1979) 239.  A. Polyakov  Gauge Fields and Strings (Harwood Academic 
Publishers, 1987) 
\bibitem{dirac} P. A. M. Dirac Phys. Rev. {\bf 74} (1948) 817.  
\bibitem{man} M. Mathur, H. S. Sharatchandra, Phys. Rev. Letts. {\bf 66} 3097 (1991),
K. Bardakci, S. Samuel Phys. Rev. {\bf D 18} 2849 (1979). 
\bibitem{emil} E. T. Akhmedov et. al. Phys. Rev. {\bf D 53} 2087 (1996). 
\bibitem{wit2} E. Witten Phys. lett. {\bf B86} (1979) 283. 
\end{thebibliography}
\end{document}